# Functional brain network architecture
# supporting the learning of social networks in humans

Steven H. Tompson[1,2], Ari E. Kahn[2,3], Emily B. Falk[4,5,6],
Jean M. Vettel[1,2,7], Danielle S. Bassett[2,8,9,10,11]

Author Note

[1]Human Sciences Campaign, U.S. Combat Capabilities Development Center Army Research Laboratory, Aberdeen, MD, USA 21005
[2]Department of Bioengineering, University of Pennsylvania, Philadelphia, PA, USA 19104
[3]Department of Neuroscience, University of Pennsylvania, Philadelphia, PA, USA 19104
[4]Annenberg School of Communication, University of Pennsylvania, Philadelphia, PA, USA 19104
[5]Department of Psychology, University of Pennsylvania, Philadelphia, PA, USA 19104
[6]Marketing Department, Wharton School, University of Pennsylvania, Philadelphia, PA, USA 19104
[7]Department of Psychological and Brain Sciences, University of California, Santa Barbara, Santa Barbara, CA, USA 93106
[8]Department of Electrical & Systems Engineering, University of Pennsylvania, Philadelphia, PA, USA 19104
[9]Department of Neurology, University of Pennsylvania, Philadelphia, PA, USA 19104
[10]Department of Physics & Astronomy, University of Pennsylvania, Philadelphia, PA, USA 19104
[11]Department of Psychiatry, University of Pennsylvania, Philadelphia, PA, USA 19104

Correspondence concerning this article should be addressed to Danielle S. Bassett, Department of Bioengineering, University of Pennsylvania, Philadelphia, PA 19104
Contact: dsb@seas.upenn.edu

**Abstract**
Most humans have the good fortune to live their lives embedded in richly structured social groups. Yet, it remains unclear how humans acquire knowledge about these social structures to successfully navigate social relationships. Here we address this knowledge gap with an interdisciplinary neuroimaging study drawing on recent advances in network science and statistical learning. Specifically, we collected BOLD MRI data while participants learned the community structure of both social and non-social networks, in order to examine whether the learning of these two types of networks was differentially associated with functional brain network topology. From the behavioral data in both tasks, we found that learners were sensitive to the community structure of the networks, as evidenced by a slower reaction time on trials transitioning between clusters than on trials transitioning within a cluster. From the neuroimaging data collected during the social network learning task, we observed that the functional connectivity of the hippocampus and temporoparietal junction was significantly greater when transitioning between clusters than when transitioning within a cluster. Furthermore, temporoparietal regions of the default mode were more strongly connected to hippocampus, somatomotor, and visual regions during the social task than during the non-social task. Collectively, our results identify neurophysiological underpinnings of social versus non-social network learning, extending our knowledge about the impact of social context on learning processes. More broadly, this work offers an empirical approach to study the learning of social network structures, which could be fruitfully extended to other participant populations, various graph architectures, and a diversity of social contexts in future studies.

## Introduction

A defining feature of modern social life, especially in Western, industrialized contexts, is the agency that individuals have in constructing their own social environment[1]. People may change jobs every few years[2], attend college far from home[3,4], move cities for work and relationships[5], and use social media to join clubs and interest groups[6]. In order to navigate these novel social environments, individuals must learn about the complex web of social relationships and cliques that characterize each new social context. For example, integration into friendship networks in the first year of college predicts future success[7] and more diverse social networks in immigrants predict better psychological well-being and cultural adjustment[8]. Understanding how people learn relational information about social networks, including which individuals each person is friends with and which communities each person belongs to, may provide key insights into how individuals adapt to novel social contexts.

Efforts to probe the learning of social relations and their complex architectures have traditionally been stymied by the lack of a formal approach for the study of network learning in general. Yet recent advances have met this challenge by conceptualizing relational information as a network in which nodes represent objects or concepts, and in which edges represent shared content or conditional probabilities[9,10]. Using tasks that are constructed based on this formal graphical conceptualization, evidence suggests that human learners are sensitive to the individual edges in a network of relational information. Perhaps even more strikingly, humans learners are also sensitive to the network's meso-scale structure, which can take on many forms including compositions of clusters or communities[11,12]. Importantly, some evidence suggests that such meso-scale structure facilitates more efficient information processing[13–15]. For example, the degree to which words are clustered together into communities is associated with how easily a particular word is learned[16]. Moreover, individuals that perform a basic perceptual learning task tend to process stimuli more slowly if the stimuli lie in different communities[11,17]. Yet, while there is extensive literature on how people learn relational information and update mental representations of language[16,18,19], motor sequences[12,20,21], and temporal associations of visual patterns[11,17,22–24], little is known about the cognitive and neural processes supporting the learning of relational information in social networks.

To address this gap, we constructed a social network learning task in which social images comprised the nodes of a modular graph, and in which relations among social images are encoded as edges linking the nodes[25]. During the task, participants were exposed to a continuous stream of social images defined by a random walk on the graph, thereby holding the transition probabilities implicit in the graph constant[22,26]. To evaluate the specificity of our results, we compared our main findings to those obtained during the learning of relational information in non-social networks. For the social portion of the experiment, participants were told that each image in a set of visual stimuli represented a person, whereas for the non-social control portion of the experiment, participants were told that each image in an equivalent set of visual stimuli represented a rock formation. Importantly, images for the social and non-social task were randomly assigned to each condition for each participant, and thus the only difference between the two tasks was the meaning ascribed to the stimuli.

The task was performed in an MRI scanner during continuous acquisition of BOLD signal. To ensure sensitivity to distributed processing and coordination across brain regions and systems, we capitalized on recently developed tools in the field of network neuroscience[27] to characterize the brain networks supporting the learning of relational information about social networks. Specifically, we identified cortical and subcortical areas that displayed strong connectivity to the rest of the brain; in the parlance of network science, these areas are referred to as hubs, and are thought to be central to coordinating communication between different brain systems and to integrating information from different systems[28]. We used whole-brain psychophysiological interaction[29] (WB-PPI; see Figure 1) to assess changes in brain networks during the task and to identify the hubs and cognitive systems that are involved in social network learning versus non-social network learning. Generally, we hypothesized that the manner in which people learn relational information about social networks, including which communities individuals belong to, should share some overlapping processing and mechanisms with how people learn information about other types of statistical relationships. Specifically, we hypothesized that brain regions involved in general memory processes such as the hippocampus[30,31] should operate as hubs supporting both social and non-social network learning. The reasoning behind our hypothesis is that the hippocampus is important for encoding relational knowledge in a variety of domains[24,32–36]. We also hypothesized that individual differences in hippocampal connectivity would be associated with individual differences in learning for both the social task and the non-social control task.

Notably, the network of hubs supporting social network learning may reconfigure relative to non-social network learning, in which case one would expect that social brain areas would operate as hubs for the social condition but not for the non-social control condition. In line with this reasoning, we hypothesized that brain regions involved in social processing such as the amygdala, dorsal medial prefrontal cortex (dmPFC), and temporoparietal junction (TPJ)[37–39] should operate as hubs preferentially supporting social network learning. Our reasoning is based on the fact that temporoparietal regions including TPJ and posterior superior temporal sulcus (pSTS) operate as hubs supporting social perception[40] and integrate sensory information with information about the broader social context to support decision-making[41,42]. We further expected that the strength of the connections between hubs would differ for social network learning and non-social network learning, such that hippocampus is more strongly connected to social hubs during the social network learning task, and more strongly connected to non-social hubs during the non-social network learning task. Finally, we hypothesized that individual differences in network learning would be associated with individual differences in functional brain network architecture, and that this relationship would be moderated by social context. Individuals who recruit social brain regions more should display enhanced performance in the social network learning task but no change in performance in the non-social control task.

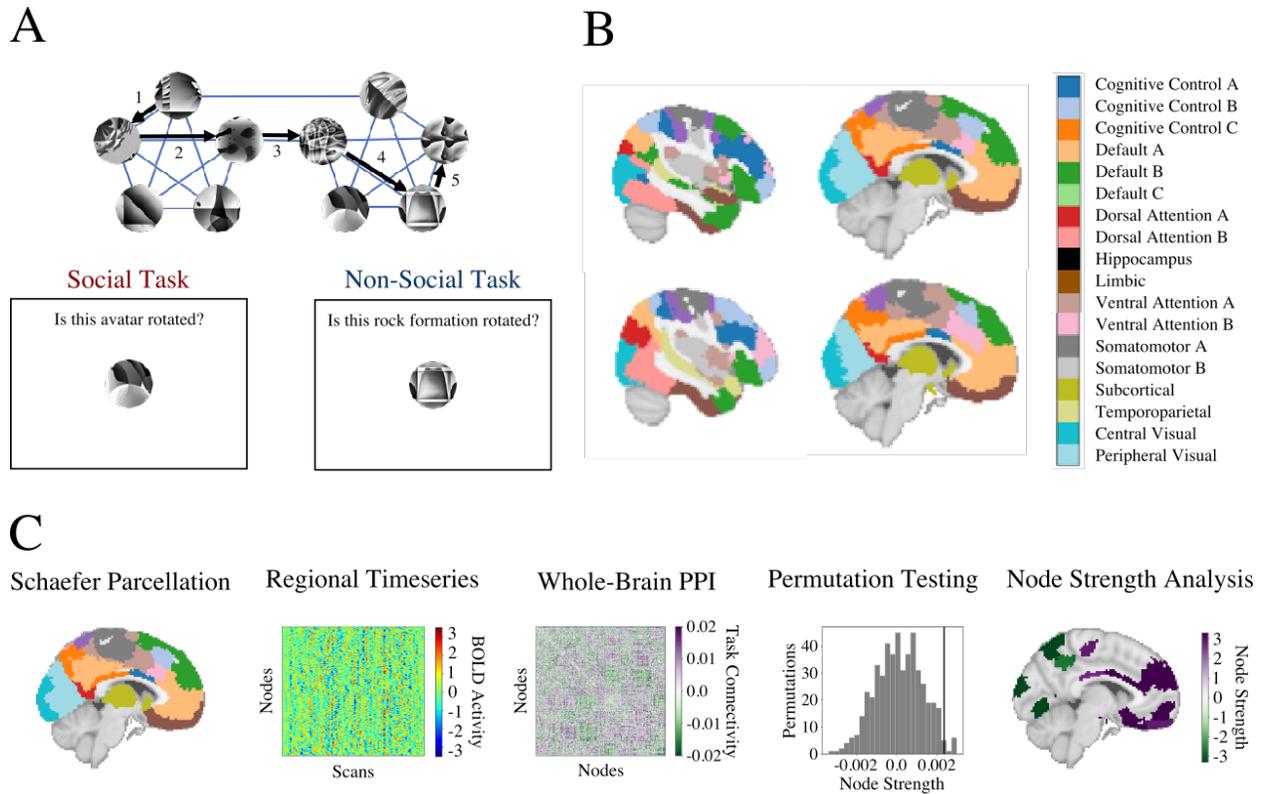

**Figure 1. Schematic of methods.** (*A*) Illustration of a random walk on a graph in which nodes are fractal images, and edges indicate allowable transitions between nodes. In the social network learning task, participants were told that the images were avatars representing people; in the control condition, participants were told that the images were rock formations. During the experiment, participants only saw the continuous stream of images, without any explicit indication of the graph from which the stream was drawn. Here we depict the graph to convey to the reader the statistical relationships among the images critical for the analysis. (*B*) Functional MRI data was collected continuously as participants performed the task. A network model of functional connectivity was constructed by first parcellating the brain into 400 cortical regions of interest using the Schaefer atlas[43] and then complementing that cortical parcellation with a parcellation of the subcortex composed of 10 additional regions of interest using the Harvard-Oxford atlas[44]. Each parcel was represented as a node in a brain network. (*C*) We used a whole-brain psychophysiological interaction model[29] to simultaneously estimate connectivity and its relation to the task. We first extracted the average timeseries from 410 brain regions defined by the Schaefer and Harvard-Oxford atlases. For each pair of regions *i* and *j*, we computed a multiple regression with the timeseries of activation in region *i* as the dependent variable and the timeseries of activation in region *j* as the independent variable; we also included a boxcar function to represent the timeseries of each task, and we used a separate interaction term to represent the interaction between activation in region *j* and the task timeseries. We constructed a 410×410 functional connectivity matrix where the $ij^{th}$ element of the matrix represented the task-dependent connectivity (beta weight for the interaction term) between region *i* and region *j*. This process was then repeated 500 times by shuffling the condition labels for the trials to generate null models of connectivity for each pair of brain regions. We computed node strength as the sum of the connectivity of each brain region with all other brain regions. We then compared the true node strength metric to the distribution of null model values to derive a *p*-value or *z*-score, with higher *z* values indicating a score that is stronger than would be expected given a random trial order.

## Results

**Behavioral evidence for network learning**

We began our investigation by assessing participant reaction time, and its dependence on the graph architecture from which the stream of stimuli were drawn. Initially, the data were collapsed across the social and non-social learning tasks, and we observed significant cross-cluster surprisal, as indicated by an increase in reaction time on trials transitioning from one cluster to another than on trials moving within a cluster ($B$=0.069, $SE$=0.014, $t$(23.644)=5.043, $p$<0.001). We also observed a significant decrease in reaction time in later trials compared to earlier trials ($B$=-0.104, $SE$=0.023, $t$(30.668)=-4.563, $p$<0.001). When we examined the two conditions separately, we found a significant cross-cluster surprisal effect for both the social network task ($B$=0.057, $SE$=0.017, $t$(23.281)=3.414, $p$=0.002) and the non-social control task ($B$=0.087, $SE$=0.024, $t$(24.569)=3.675, $p$=0.001), consistent with the idea that participants were able to learn both graph structures. Notably, participants did not differ in their mean accuracy for the social network task ($M$=84.8%, $SD$=10.1%) and for the non-social control task ($M$=84.9%, $SD$=9.3%, $t$(25)=0.091, $p$=0.928).

**Functional hubs that support network learning**

We next sought to identify the brain network features that support network learning generally. We examined the relationship between brain networks and network learning at two levels of resolution: the region-level and the system-level. This approach enables us to examine both which specific brain regions are involved in network learning as well as which groups of brain regions are involved in network learning.

We began by identifying the functional hubs that were shared across the social network task and the non-social control task. Specifically, we computed the average strength for each node across the brain networks extracted by PPI during both conditions. We normalized each strength value to obtain a $z$-score, by comparing the node strength to a null distribution (see Methods and Figure 2A). We found that hippocampus, thalamus, dmPFC, STG, temporal pole, and OFC displayed significantly greater node strength for transition trials than for non-transition trials across both tasks (FDR corrected $p$<0.05 over brain regions; Figure 2B). In contrast, visual cortex, IPL, TPJ, insula, and dlPFC exhibited significantly greater node strength for non-transition trials than for transition trials across both tasks (FDR corrected $p$<0.05 over brain regions; the full list of regions is included in Table S1 in Supplementary Results).

To better understand this pattern of results, we assessed to what degree functional hubs were found within putative functional systems defined by an *a priori* community assignment[44,45]. We found that 86% of the regions displaying significantly greater node strength for transition than non-transition trials were concentrated in hippocampus, limbic system, default mode systems, and sensory systems (see Figure 2C). In particular, the number of such regions in the hippocampus ($Z$=2.748, FDR-corrected $p$=0.027) and limbic system ($Z$=3.291, FDR-corrected $p$<0.001) were significantly greater than expected in a non-parametric permutation-based null model in which regions were randomly assigned to systems. In contrast, we found that 78% of the regions displaying significantly greater node strength for non-transition than transition trials were concentrated in cognitive control and attention systems. In particular, the number of hubs in the cognitive control subsystem C ($Z$=2.878, FDR-corrected $p$=0.018), the dorsal attention subsystem A (*Z*=3.291, FDR-corrected *p*<0.001), and the dorsal attention subsystem B (*Z*=2.226, FDR-corrected *p*=0.039) were significantly greater than expected in the non-parametric permutation-based null model. These results indicate that different brain systems may underlie the ability to learn different features of network structures.

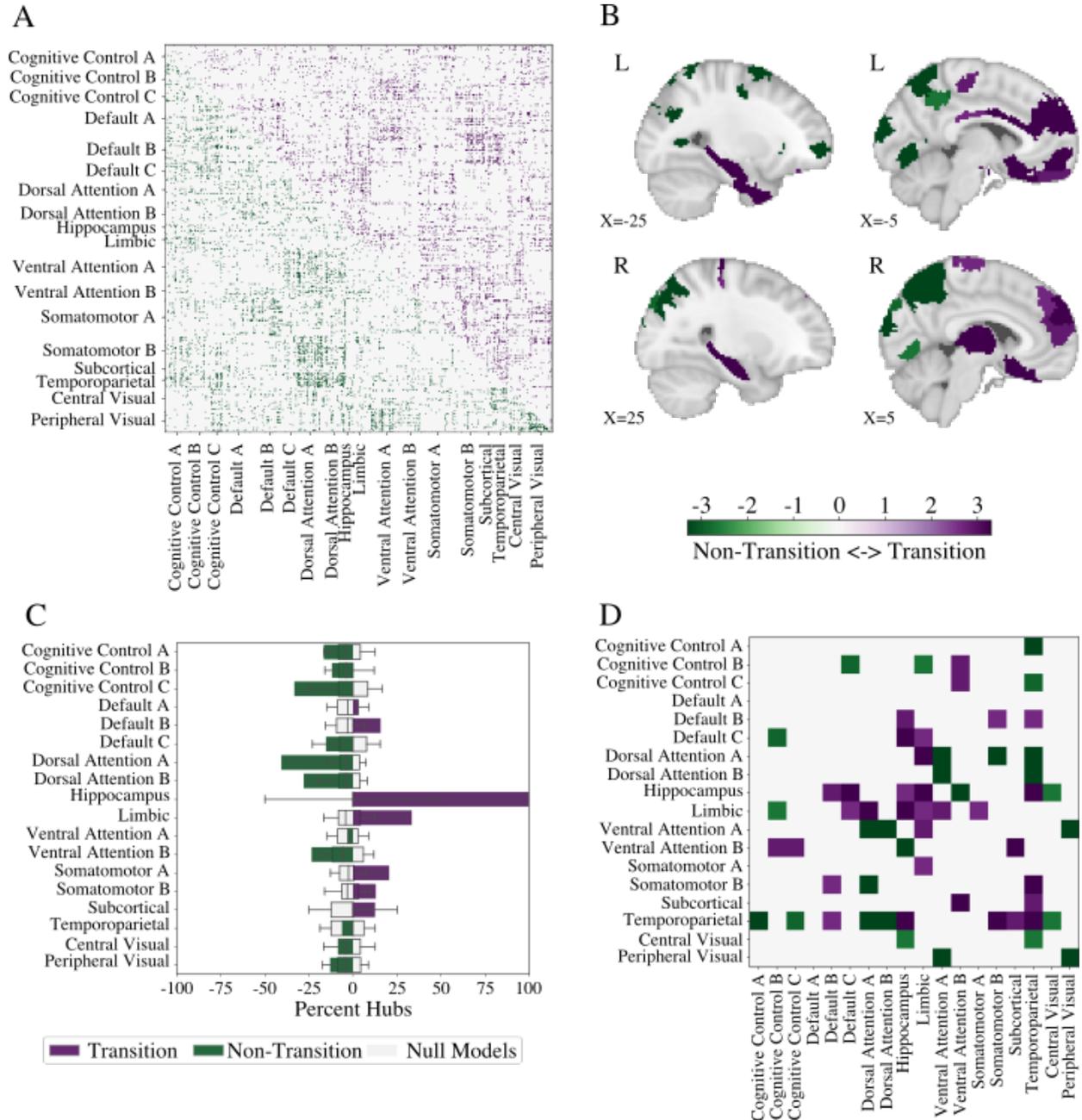

**Figure 2. Patterns of functional connectivity in the brain are distinct in transition trials compared to non-transition trials.** (*A*) Functional connectivity matrix showing edges with significantly different weights (FDR corrected *p*<0.05 over edges) for transition (upper triangle in purple) versus non-transition (lower triangle in green) trials. (*B*) Whole-brain map showing regions with significantly different strength *z*-scores for transition (purple) versus non-transition (green) trials (FDR corrected *p*<0.05 over brain regions). (*C*) Bar graph showing the percentage of network hubs identified in each cognitive system relative to the percentage identified in a non-parametric null model (gray box plots). (*D*) By averaging edge weights both within and between putative functional

systems, we constructed a system-level connectivity matrix. Here we display that matrix showing systems with significantly different connectivity for transition (purple) versus non-transition (green) trials (FDR corrected $p<0.05$ over edges in the system-level matrix).

In a complementary analysis, we examined these relationships at a coarser, system level by averaging edge weights within and between the same putative functional systems[44,45], and then we tested whether these coarse-grained elements differed for transition and non-transition trials. We observed that functional connectivity within the limbic and default mode systems was significantly greater for transition trials than for non-transition trials. Furthermore, functional connectivity between the hippocampus and default mode system, between the hippocampus and somatomotor system, and between the default mode and somatomotor systems, was significantly greater for transition trials than for non-transition trials. In contrast, connectivity within the visual system and connectivity between frontal cognitive control systems and default mode systems was significantly greater for non-transition trials than for transition trials (see Figure 2D). Taken together, these results suggest that memory, default mode, and sensory systems have greater node strength as well as connectivity with each other during transitions between clusters than during trials occurring within a cluster.

**Functional hubs that preferentially support social network learning**

After characterizing functional hubs involved in both conditions, we next sought to determine whether some hubs preferentially supported social network learning. We once again examined the relationship between brain networks and network learning at a coarse-grained systems level as well as a more fine-grained region level, which allows us to examine which specific brain regions are preferentially involved in social network learning as well as which groups of brain regions are preferentially involved in social network learning.

To address this question, we computed the difference in average node strength for each potential hub for social versus non-social networks. We normalized each difference value to obtain a $z$-score by comparing it to a null distribution (see Methods). We observed significantly greater node strength for the social network condition than the non-social control condition in TPJ, PCC, middle cingulate, precentral gyrus, postcentral gyrus, and visual cortex (FDR-corrected $p<0.05$; see Figure 3B; full list of regions is included in Table S2 in Supplementary Results). For completeness, we also assessed whether some hubs were stronger in the non-social control condition. We observed significantly greater node strength in the non-social control condition than in the social network condition in bilateral dmPFC, IFG, middle cingulate, inferior temporal lobe, visual cortex, and dlPFC (FDR-corrected $p<0.05$; see Figure 3B and Table S2). Collectively, these results suggest distinct regional involvement in social network learning in comparison to a non-social control condition.

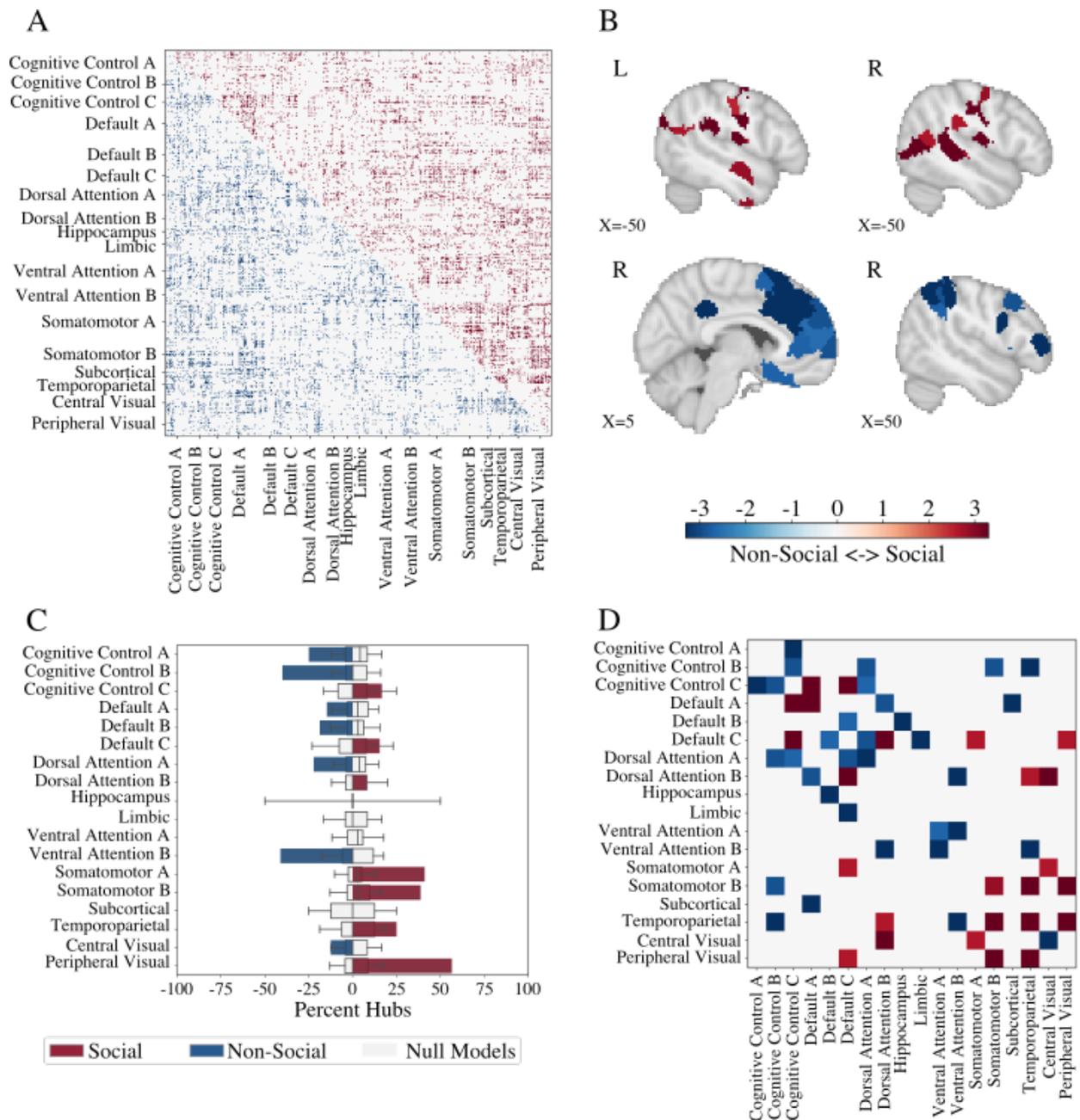

**Figure 3. Patterns of functional connectivity in the brain are distinct for social network learning compared to a non-social control condition.** (*A*) Functional connectivity matrix showing edges with significantly different weights for the social network condition (upper triangle in red) versus the non-social control condition (lower triangle in blue). Significance was assessed after an FDR correction at $p<0.05$ over edges. (*B*) Whole-brain map showing regions with significantly different strength $z$-scores for trials in the social network condition (red) versus in the non-social control condition (blue). Significance was assessed after an FDR correction at $p<0.05$ over brain regions. (*C*) Bar graph showing percentage of network hubs identified in each cognitive system for the social network condition versus the non-social control condition, relative to the percentage identified in a non-parametric null model (gray box plots). (*D*) By averaging edge weights both within and between putative functional systems, we constructed a system-level connectivity matrix. Here we display that matrix showing systems with significantly different connectivity for the social network condition (red) versus the non-social control condition (blue). Significance was assessed after FDR correction at $p<0.05$ over edges in the system-level matrix.

To better understand this pattern of results, we assessed to what degree these functional hubs were found within putative functional systems defined by an *a priori* system assignment[45]. We found that 81% of the regions displaying significantly greater node strength in the social network condition than in the non-social control condition were concentrated in the default mode, somatomotor, and visual systems (Figure 3D). In particular, the number of such regions in somatomotor subsystem A (*Z*=3.291, FDR-corrected *p*<0.001), somatomotor subsystem B (*Z*=3.291, FDR-corrected *p*<0.001), and peripheral visual subsystem (*Z*=3.291, FDR-corrected *p*<0.001) were significantly greater than expected in a non-parametric permutation-based null model in which regions were assigned to systems uniformly at random. The number of such regions in the temporoparietal default subsystem trended towards significance after FDR-correction (*Z*=1.995, FDR-corrected *p*=0.090). In contrast, we found that 65% of the regions displaying significantly greater node strength for the non-social control condition than the social network condition were concentrated in cognitive control and attention systems. In particular, the number of such regions in cognitive control subsystem A (*Z*=3.291, FDR-corrected *p*<0.001), cognitive control subsystem B (*Z*=3.291, FDR-corrected *p*<0.001), default subsystem B (*Z*=2.290, FDR-corrected *p*=0.050), dorsal attention subsystem A (*Z*=2.290, FDR-corrected *p*=0.050), and ventral attention subsystem B (*Z*=3.291, FDR-corrected *p*<0.001) were significantly greater than expected in the same non-parametric permutation-based null model. The number of such regions in default subsystem A was marginally significant after FDR-correction (*Z*=1.995, FDR-corrected *p*=0.090).

In a complementary analysis, we examined these relationships at a coarser, system level by averaging edge weights within and between the same putative functional systems[45], and then we tested whether these coarse-grained elements differed for the social network condition compared to the non-social control condition. We observed that functional connectivity was significantly greater within the default mode and somatomotor systems for the social network condition than for the non-social control condition (FDR-corrected p<0.05; see Figure 3E). In contrast, we observed that functional connectivity was significantly greater within the dorsal attention, ventral attention, and central visual systems for the non-social control condition than for the social network condition (FDR-corrected p<0.05). Expanding our assessment to both within- and between-system connectivity, we found that the temporoparietal, default mode, and somatomotor systems were more strongly connected to each other and to the peripheral visual subsystem in the social network condition, suggesting that the hubs that we identified in the temporoparietal cortex may be integrating sensory information with the social context. In contrast, the non-social control condition was primarily characterized by increased connectivity between frontal cognitive control and attention systems, as well as limbic systems and frontal default systems. Thus, cognitive systems involved in social network learning reconfigure to form distinct subnetworks from those involved in non-social network learning.

**Specific connectivity patterns of functional hubs support social network learning**

To complement the whole-brain analyses reported in the previous sections, we wished to evaluate specific connectivity patterns emanating from functional hubs that might preferentially support social network learning. We began our investigation by considering the specific role of the hippocampus, largely motivated by important prior work offering evidence in favor of its role in similar tasks[32,33]. We found that the hippocampus was connected to different regions during the social network task compared to the non-social control task (Tables S4 and S5 in

Supplementary Results). In particular, the left hippocampus had significantly stronger connectivity with the dmPFC and IFG during the non-social control condition (FDR-corrected *p*<0.05; Figure 4A), whereas the left hippocampus had significantly stronger connectivity with TPJ during the social network condition than the non-social control condition (FDR-corrected *p*<0.05). The right hippocampus also had significantly stronger connectivity with IFG during the non-social control condition (FDR-corrected *p*<0.05) and significantly stronger connectivity with bilateral TPJ during the social network condition (FDR-corrected *p*<0.05). During both conditions, we found that the hippocampus exhibited significantly greater connectivity with hubs specific to the non-social control task, including the dmPFC and IFG (FDR-corrected p<0.05).

To examine if these relationships existed at the systems-level, we averaged edge weights of hubs within *a priori* regions of interest (hippocampus, dmPFC, lPFC, and TPJ) in order to examine how connectivity between hubs involved in both conditions might interact with hubs preferentially supporting learning in the social network task or the non-social control task. We found that hippocampus exhibited significantly greater connectivity with non-social hubs including dmPFC and IFG for non-social tasks than social tasks (FDR-corrected *p*<0.05; Figure 4B). This set of findings suggests that brain regions that have the strongest connectivity during non-social (relative to social) tasks are engaging with memory systems in support of learning the network structure of non-social networks.

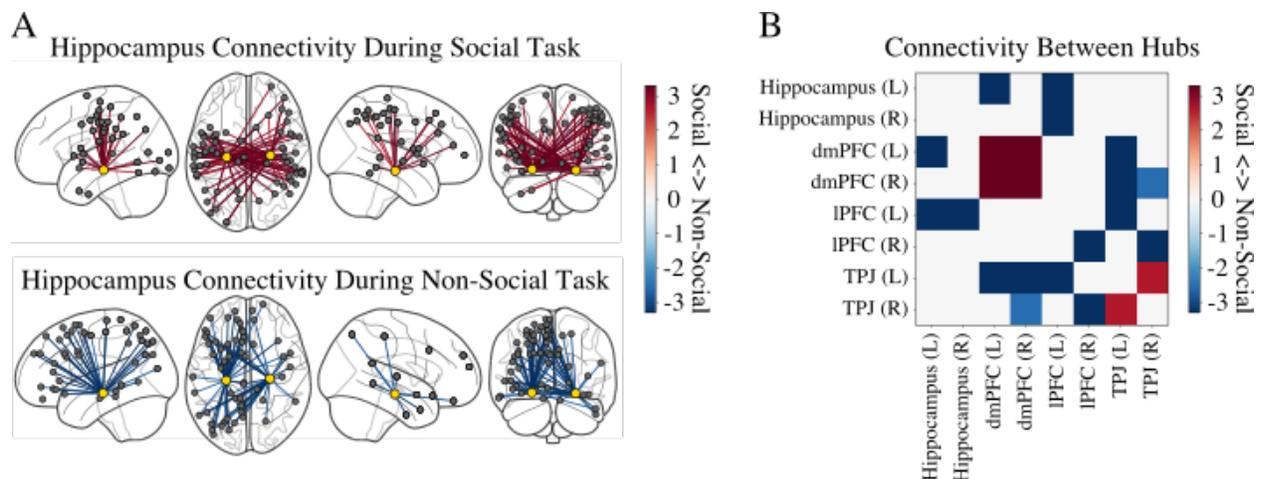

**Figure 4. Functional connectivity between the hippocampus and other brain areas.** (*A, top*) Significant edges representing greater connectivity between the hippocampus (indicated in yellow) and other hubs during the social network learning condition than during the non-social control condition (top image). (*A, bottom*) Significant edges representing greater connectivity between the hippocampus (indicated in yellow) and other hubs during the non-social control condition than during the social network condition (bottom image). Significance is assessed after FDR correction at *p*<0.05 over the number of edges. (*B*) Connectivity matrix representing the average connectivity within and between hubs in *a priori* defined ROIs, thresholded at an FDR-corrected *p*<0.05.

**Relation between brain connectivity and behavioral performance**

Finally, we sought to test whether individual differences in hub connectivity were associated with how well individuals learned the network architecture from which the sequence of stimuli were drawn. We extracted the average node strength for each participant in the hubs that we identified above as being located in *a priori* brain regions (TPJ, PFC, hippocampus). Specifically, we considered (i) social hubs as the TPJ regions that had significantly greater node strength for the social network condition than for the non-social control condition, (ii) non-social hubs as the medial and lateral PFC regions that had significantly greater node strength for the non-social control condition than for the social network condition, and (iii) domain-general hubs as hippocampus, TPJ, and mPFC regions that had significantly greater node strength than other brain areas across both conditions. Participants with greater node strength in social hubs displayed a significantly stronger cross-cluster surprisal effect for the social network condition but not for the non-social control condition ($r(24)=0.478$, $p=0.013$, $r(24)=0.168$, $p=0.411$, respectively; see Figure 5A). Importantly, this effect was driven by connectivity between social hubs in the left and right TPJ, such that the average connectivity between social hubs in left and right TPJ was significantly correlated with cross-cluster surprisal in the social network condition but not in the non-social control condition ($r(24)=0.446$, $p=0.022$, $r(24)=-0.087$, $p=0.674$, respectively; see Figure 5B). We observed no significant relationships between cross-cluster surprisal and the non-social hubs or the domain-general hubs.

We next tested whether the relationship between hub connectivity and cross-cluster surprisal was significantly different for social versus non-social network learning. To address this question, we constructed a linear mixed effects model with node strength and network type (social versus non-social) as predictor variables and cross-cluster surprisal as the dependent variable. For social hub node strength, we observed a significant main effect of node strength ($B=0.034$, $SE=0.015$, $t(47.47)=2.313$, $p=0.025$). However, the main effect of network type ($B=0.026$, $SE=0.029$, $t(26.12)=0.899$, $p=0.377$), and the interaction between node strength and network type ($B=0.02$, $SE=0.015$, $t(47.11)=1.333$, $p=0.189$) were not significant. For the strength of connectivity between social hubs in left and right TPJ, we observed a marginally significant interaction between the strength of connectivity and network type ($B=0.029$, $SE=0.015$, $t(44.38)=1.983$, $p=0.054$), but no main effect of strength of connectivity ($B=0.024$, $SE=0.015$, $t(47.54)=1.564$, $p=0.124$). or main effect of network type ($B=0.026$, $SE=0.027$, $t(24.64)=0.958$, $p=0.347$). Thus, social hubs in TPJ appear to be involved in the learning of social network structure, in part due to their increased connectivity with one another, although the relationship is only marginally stronger for social versus non-social networks.

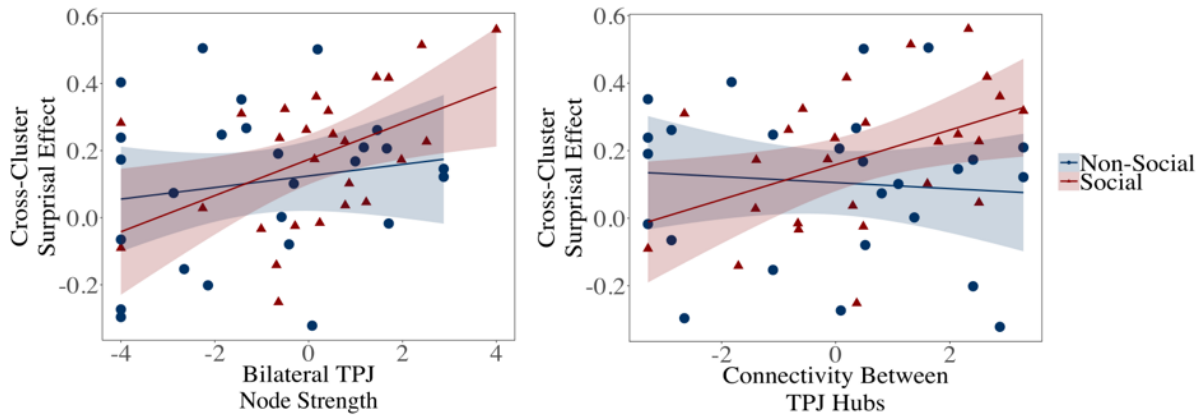

**Figure 5. Association between brain connectivity and network learning.** (A) Participants who had higher node strength scores in bilateral TPJ showed stronger cross-cluster surprisal for the social network learning condition, but not for the non-social control condition. (B) Participants who had stronger connectivity between social hubs in the TPJ also showed stronger cross-cluster surprisal for the social network learning condition, but not for the non-social control condition.

## Discussion

In this paper, we investigate the brain networks involved in the implicit learning of community structure for social networks. Navigating interwoven layers of social connections is critical for success in a broad range of social interactions with co-workers, friends, family and strangers[46–49]. Although there is an extensive literature examining how people learn relational information in the domains of language learning[16,19], motor sequence learning[12,20,21], and statistical learning[11,17,26], little is known about the neural mechanisms underlying how people learn relational information about social networks. Building on this extant literature, we showed both overlap and divergence in the brain networks involved in social and non-social network learning. We found that the hippocampus operated as a hub supporting both types of network learning. Temporoparietal regions of the default mode system operated as hubs preferentially supporting social network learning, where they were also more strongly connected to the hippocampus as well as to somatomotor and visual systems. Medial and lateral prefrontal cortical regions in both the default mode system and frontal cognitive control system operated as hubs preferentially supporting non-social network learning, where they were also more strongly connected to the hippocampus as well as to cognitive control and attention systems. Furthermore, individuals who had stronger hubs in the temporoparietal default system were better at learning the community structure of the social networks. This work extends our understanding of how brain networks and social context are associated with learning. In particular, this work provides insight into how individuals learn about features of social networks and how the brain reconfigures to support this learning process.

**Memory systems support learning both social and non-social networks**

Consistent with prior work examining how people learn relational information in other domains[20], the hippocampus and PFC have also been found to play an integral role in the learning of community structure present in networks comprised of visual images[11,33]. Evidence suggests that multivariate patterns in the hippocampus and IFG represent information about which communities individual nodes belong to, whereas mPFC represents boundaries and transitions[11,33]. Importantly, lesion studies also suggest that the hippocampus is necessary for learning the statistical relationships between non-social objects [32,50]. Here, we extend this work by applying graph theory to show that the hippocampus is a primary hub supporting both social and non-social network learning. Both the hippocampus and parahippocampal gyrus have significantly greater node strength for transition trials than for non-transition trials. This pattern of results is consistent with work suggesting that the hippocampus integrates representations of objects and their spatiotemporal context from disparate cortical areas,[51] as well as evidence suggesting that hippocampal-cortical connections support simulation of future navigation in complex spatial maps[34].

At the system-level, the hippocampus exhibits stronger connectivity with the default mode system for transition trials than for non-transition trials. This pattern of results is consistent with prior studies providing evidence that the default mode system is involved in developing and maintaining models that predict and simulate future events[52]. In the context of learning community structure in networks, transition trials may be interpreted as prediction errors, given that participants are responding more slowly because they anticipated the next trial to remain within a community[11,12,53]. In the context of spatial navigation, some researchers have argued that prefrontal cortex monitors prediction errors and updates contextual information about the prediction, which is then integrated by the hippocampus into an updated mental model[54]. Similarly, medial default mode network regions process temporal information about motor patterns and support the testing of predictive models of future motor sequences[55]. It is intuitively plausible that the hippocampus and regions of the default mode may be communicating with one another to update predictions about future trials and representations of the community structure.

Although the hippocampus operates as a hub during both social and non-social network learning, the pattern of connections to cortical areas differs for the two conditions. At the region level, the hippocampus is primarily connected to temporoparietal, somatomotor, and visual areas during social network learning, whereas it is primarily connected to medial prefrontal, lateral prefrontal, and inferior temporal areas during non-social network learning. This differentiation of connectivity is consistent with past work showing that temporoparietal regions support updating of prediction errors by validating mental models against sensory inputs[56]. Collectively, the pattern of findings suggests that there exist marked differences in how the hippocampus is integrating information to update representations of the community structure of social versus non-social networks.

**Nexus hubs versus top-down control of network learning**

Beyond the role of memory systems in network learning, we observed notable connectivity patterns in temporoparietal brain areas including the right TPJ. Temporoparietal, default, somatomotor, and peripheral visual subsystems were more strongly connected to each other during social network learning than during the non-social control condition. This pattern of results is consistent with prior work that suggests that TPJ may integrate sensory information with information about the broader social context to support decision-making[40–42], in part due to its location at the intersection of brain regions involved in many different types of cognitive processes[42]. In light of these prior studies, it would appear possible that TPJ is integrating information about the visual features of the images with motor responses from the task and with the social/online avatar framing of the task to support the participants' learning. Importantly, the node strength of left and right TPJ hubs as well as the connectivity among TPJ hubs was significantly correlated with cross-cluster surprisal for the social network condition but not for the non-social control condition, supporting the region's role in social network learning specifically.

By contrast, non-social network learning primarily involved connectivity with prefrontal brain areas including the dmPFC and IFG. Frontal cognitive control and attention subsystems were more strongly connected to each other, as well as more strongly connected to medial prefrontal default subsystems, during non-social network learning than during social network learning. One possible interpretation of this result is that cognitive control and attention areas are exerting top-down control to direct processing of sensory information[54,57,58] in support of developing a mental representation of the non-social network. Here, prefrontal cortical regions may be either modulating sensory processing of stimuli to support the goal of learning the non-social network[58], or may be modulating attention to information that hippocampus will integrate into an updated mental representation[54].

It is also interesting to note that we did not find individual differences in any brain hubs or systems that were significantly correlated with individual differences in the magnitude of the cross-cluster surprisal effect during the non-social control condition. Contrasting the group-level findings with these individual-level findings, our data suggest that prefrontal systems are consistently engaged in the cognitive processes that we study here, but that individual differences in behavior may require explanations that include other regional features, or other brain regions altogether.

**Representing features of real-world social networks**

Our study expands upon recent work investigating how people represent the positions of individuals in a social network. Recent work has found that people automatically track and encode information about the position of peers in their immediate social network. In brain regions associated with mentalizing and reward, college students represent information about the social value of other students in their dormitory[59]. People also display neural representations of information about others' social position including their network centrality, brokerage opportunities in the network, and social distance from the perceiver[60–62]. Collectively, these studies focus on the position of individual nodes in a social network, and we expand upon this work by considering meso-scale features of the social network including its community structure. We find that many of the brain regions implicated in representing the network position of individual nodes are also important for learning community structure (e.g., PCC and right TPJ). Importantly, we also show that functional connectivity between these brain regions, and functional connectivity with memory regions including the hippocampus, is important for learning meso-scale network features.

Furthermore, recruitment of TPJ has been linked to individual's position in their social network[63,64]. For example, people who occupy more central positions in their social network tend to recruit TPJ more when making sense of social information[64]. People with less dense social networks also show greater functional connectivity between left and right TPJ when making sense of social contexts such as being excluded[63] and those who are more receptive to peer influence show higher node strength of TPJ[65]. Our work expands on this extant literature by showing that people who have high TPJ node strength and greater functional connectivity between left and right TPJ learn social networks better. Learning information about the community structure of social networks might be important for how individuals position themselves in their social network, or the positions of individuals within their social network might influence which brain regions they use to process social information.

**Methodological considerations and limitations**

Several methodological considerations are pertinent to this work. First, given the contrast between the two conditions in the current paradigm, one might anticipate that brain regions involved in considering the mental states of others might be most strongly involved in the social network condition. While we do find that TPJ is more strongly involved in the social condition, other mentalizing regions including dmPFC were actually more strongly involved in the non-social control condition. Thus, it is difficult to conclude that social network learning demands additional mentalizing processes relative to non-social network learning. In the current study we used the same abstract shapes to represent people (social condition) or rock formations (non-social control condition). It is possible that a stronger manipulation, perhaps using real people or real social networks, might yield stronger recruitment of mentalizing brain regions.

Second, the experimental paradigm that we employed requires dozens of trials per node to detect the cross-cluster surprisal effect, and we were therefore limited in the size and complexity of the social network that we could test. Real-world social networks are often much larger than those we study here, and they also tend to display much more complex patterns of connections[66]. There is some evidence that learnability of network structures is robust to the size of the communities within the network[67]. But, other work on network learning in non-social domains suggests that whether a network is organized into clusters versus more random networks or networks with a lattice-like structure influences how well it is learned[12]. Future work could investigate how people learn other, more naturalistic, configurations of social networks.

Third, one limitation of past neuroimaging work on how people learn relational information is that these studies have focused on activation in single brain regions or connectivity between pairs of brain regions[11,19,20,33]. Implicitly learning the edges and community structure of a complex network and building a mental representation of that network likely involves processing and coordination from many brain regions and systems[33]. In supplementary analyses we did not find any significant differences in univariate brain activation for social versus non-social network learning (see Supplementary Results). These facts underscore the need for an explicitly multivariate approach that takes into account the distributed set of brain regions, and their interconnections, that support network learning. Here we address this need by capitalizing on recent advances in the nascent field of network neuroscience, which applies network science tools to understand how groups of brain regions interact to support both basic and complex cognitive processes across different temporal and spatial scales [27]. Specifically, we examine network hubs, or brain areas that have strong connectivity to the rest of the brain, because they are thought to be central to coordinating communication between different brain systems and integrating information from different systems[28].

**Conclusion**

In this study, we apply graph theory and network neuroscience tools to investigate how people learn information about social networks, including which communities each person in a social network belongs to. Navigating novel social contexts is a defining feature of modern social life[1] and learning information about how people in a new social group are connected to each other is an important facet of adapting to and fitting into these new social contexts[46–49]. We found that people learned the community structure of both social and non-social networks, but the learning of these two types of networks was differentially associated with brain network architecture. Brain regions implicated in memory processes operate as hubs supporting learning community structure of both social and non-social networks, whereas brain regions implicated in social processing operate as hubs supporting learning of social (but not non-social) community structure. Evidence supports the notion that social hubs integrate information from sensory systems, whereas non-social hubs in cognitive control systems are exerting top-down control of information processing. Broadly, our study provides a promising approach to determine how the brain supports social network learning, extending our knowledge about the impact of functional brain networks and social context on these learning processes.

**Methods**

**Participants**

Thirty-two participants (12 male and 20 female) were recruited from the general population of Philadelphia, Pennsylvania USA. All participants were between the ages of 18 and 65 years ($M=25.20$, $SD=9.66$), were right-handed, and met standard MRI safety criteria. Two participants opted out of the study during the scanning session due to claustrophobia, two additional participants were excluded due to an average accuracy below 50%, and two participants were excluded due to artifact that resulted in large signal dropout in the parietal lobe.

**Procedure**

While their brain activation was measured in an MRI scanner, participants completed a social network learning task and a non-social control task. The structure of the experiment was identical to Study 5 in Tompson *et al.* (2018). The order of the two tasks was counterbalanced across participants, and when we included task order as a covariate in our analyses, the results reported below were unchanged. In each task, participants viewed a sequence of fractal images that we created using the Qbist filter in the GNU Image Manipulation program (v.2.8.14; www.gimp.org), converted to grayscale, and then matched for average brightness. Images were presented for 1500 ms. To ensure that participants were attending to the stream of images, they were instructed to press the J key with their right index finger if the image was rotated (30% of trials) and to press the F key with their left index finger if the image was not rotated (70% of trials). Each task was broken into 5 runs and participants were given a break between runs to reduce fatigue.

Each image was unique, and for each participant, each image was randomly assigned to a network node in either the social or non-social condition. The sequence of fractal images that each participant saw for each task was generated by a random walk through the network (see Figure 1A). This random walk ensured that the probability of one image being presented after the current trial was equivalent across trials and determined by the network structure. Each node was connected to exactly four other nodes, ensuring that all transition probabilities were equivalent. The structure of transition probabilities is an important cue that signals event structure, which can influence how quickly participants learn information[22,26,68]. Therefore, keeping the network structure uniform to remove transition probabilities as a potential source of information about which trials to expect next is important for testing whether participants can learn higher-order network topology.

Participants completed a brief training procedure prior to starting each task. First, they were shown each image in its non-rotated orientation. Then, they were shown the rotated and non-rotated versions side by side and asked to pick the non-rotated image. Next, they completed a practice version of the rotation detection task, where they saw each image once in random order. During each task, participants were also given audio feedback to assist them in learning the rotation of images. Specifically, they heard a high tone when they made an incorrect response and a low tone when they responded too slowly (greater than 1500 ms).

The network structure for each task consisted of two clusters each composed of five nodes, and participants viewed a sequence of 1000 fractal images in each task. For the social task, participants were told that "the images that you will see are taken from an online social media platform where people can choose one of these images as their avatar to represent themselves, much like you might use a photo to represent yourself on Facebook or Twitter. While completing the task (described in more detail on the next page), please make sure you focus on the people these avatars represent." In the non-social condition, participants were told that the "images were abstract patterns frequently found in rock formations. Some of these patterns are visible to the naked eye, whereas others are only visible with a microscope. These rock patterns are often created by natural forces, including tectonic plate shifts, wind and water erosion, and volcanic activity. While completing the task (described in more detail on the next page), please make sure you focus on the patterns in the rock formations."

**Cross-cluster surprisal**

In order to estimate how well participants learned each network, we examined the cross-cluster surprisal, which is measured as the difference in RT between *pre-transition trials* that occurred immediately before a transition from one cluster to another and *post-transition trials* that occurred immediately after a transition from one cluster to another[17]. If participants learn the cluster membership, then they should anticipate seeing a within-cluster image rather than an image from another cluster[12,17,25,67]. This surprisal effect should slow participants' response to the rotation judgment on the next trial[11,17].

To examine group-level effects of the community structure on participants' responses, we used linear mixed effects models implemented with the *lmer()* function (library lme4, v. 1.1-10) in R (v. 3.2.2; R Development Core Team, 2015). The primary mixed effects model included node type (pre-transition versus post-transition), network type (social versus non-social), order (social network first versus non-social network first), trial number (standardized), and the two-way and three-way interactions between these variables, as predictors of RT (with node type, network type, and trial number included as within-subjects variables and order included as a between-subjects variable). For all models, all predictors were mean-centered and we included the fullest set of random effects that allowed the model to converge, which included a random intercept for participant and a by-participant random slope for trial number, network type, and node type. We then conducted simple effects analyses to examine whether the effect of node type was significant in both the social and non-social task conditions. To examine individual differences in participants' ability to learn the community structure, we also calculated the average cross-cluster surprisal effect for the social task and the average cross-cluster surprisal effect for the non-social task for each participant.

**fMRI Acquisition**

Data were acquired on a 3T Siemens Prisma scanner with a 64-channel head/neck array. Functional images were acquired using a T2*-weighted image sequence with a repetition time (TR) of 1000ms, an echo time (TE) of 32ms, a flip angle of 60°, and a 20cm FOV consisting of 56 with 2.5mm thickness acquired at a negative 30° tilt to the AC-PC axis, with a 2.5mm isotropic voxel size, and a multiband factor of four. We also acquired a high-resolution structural image using a T1-weighted axial MPRAGE sequence yielding 160 slices with a 0.9 by 0.9 by 1.0mm voxel size.

**fMRI Data Preprocessing**

Neuroimaging data was preprocessed using nipype[69] and nilearn[70] implemented in python 2.7 using a combination of AFNI[71], FSL[72,73], and ANTs[74]. The functional data underwent de-spiking to smooth outliers in each voxel using AFNI's 3dDespike, rigid transformation to correct for head motion using FSL's MCFLIRT[75], and slice-time correction using FSL's Slicetimer to control for temporal differences in the order of the acquisition of the slices in each brain volume. Skull stripping was performed on the structural data using FSL's BET. The mean functional image was computed and bias-corrected, and then the skull-stripped structural image was bias-corrected. Advanced Normalization Tools (ANTs) was used to compute the transformation parameters for the mean functional image to the high resolution structural image. ANTs segmentation was performed to obtain a warped structural image, a skull-stripped brain mask, and masks for white matter and cerebrospinal fluid. Confound regression was then conducted. The time series was detrended by regressing the time series on the mean and polynomial trends up to quadratic terms. AFNI's 3dbandpass was used to filter out very high or very low fluctuations in the signal (with a high pass of 0.01 and a low pass of 0.12). We included 36 regressors in the confound regression, with six head-motion regressors, three physiological signal regressors (global signal, white matter, and cerebrospinal fluid), as well as their derivatives, quadratics, and squared derivatives. ANTs was used to warp the high-resolution structural image to the MNI template. The transformation parameters from the ANTs functional to structural co-registration and the transformation parameters from the ANTs structural to MNI co-registration were used to warp the 4D functional image to the MNI template. Finally, high variance compounds were removed[76] using nilearn. The current preprocessing stream was chosen based on studies that evaluated the performance of a wide variety of preprocessing pipelines in mitigating motion artifact in studies of BOLD functional connectivity[77,78].

**Functional Connectivity**

In order to examine the functional brain network architecture of network learning, we used the local-global Schaefer cortical parcellation[43] that divides the human cerebral cortex into 400 functionally homogenous regions (Figure 1B). Given our interest in memory, learning, and social processes, we also added 10 subcortical regions in the left and right hippocampus, left and right amygdala, left and right ventral striatum, left and right caudate, and left and right thalamus using the Harvard-Oxford subcortical atlas[44].

To assess functional connectivity between the regions identified above, we first extracted the average timeseries of activation in each region and standardized the timecourse in each region using the nilearn[70] package in Python 2.7. We then computed the task-dependent connectivity between each pair of regions using a whole-brain psychophysiological interaction (WB-PPI) approach[29]. For each pair of regions $i$ and $j$, we computed a multiple regression with the timeseries of activation in region $i$ as the dependent variable and the timeseries of activation in region $j$ as an independent variable; we also included a boxcar function to represent the timeseries of each task (coded as 1 during transition trials and 0 during non-transition trials), and we used a separate interaction term to represent the interaction between activation in region $j$ and the task timeseries. In order to compute the interaction term, the timeseries in region $j$ was first deconvolved from the canonical HRF function and then multiplied by the boxcar function. The boxcar function and interaction term were then re-convolved with the canonical HRF before computing the multiple regression model. Six head motion parameters and a constant term were also included in the model. These processes were implemented in Python 2.7 using a combination of functions from the nilearn, nipy, and nistats packages, and were designed to follow as closely as possible the implementation of generalized PPI in SPM[79]. We constructed a 410×410 functional connectivity matrix where the $ij^{th}$ element of the matrix represented the task-dependent connectivity (beta weight for the interaction term) between region $i$ and region $j$. We then symmetrized the matrices for each run by averaging the upper and lower triangles, and then we averaged the functional connectivity matrices for the five runs for each task for each subject to yield two functional connectivity matrices for each subject (one for the social task and one for the non-social task).

Given that the transition trials were determined by a random walk through the graph, there were a varying number of transition trials across runs and subjects. In order to account for this variation, we adopted a nonparametric permutation test of significance. We constructed 500 null model networks for each subject by shuffling the trial labels and rerunning the PPI analyses described above, keeping the number of transition trials in each run consistent. The true PPI beta weight for each edge (or other summary statistic) can be compared to the distribution of null model values to derive a *p*-value or *z*-score. Higher positive *z*-scores represent stronger functional connectivity in transition trials versus non-transition trials than would be expected given a random trial order.

**Hub Analysis**

To identify which brain regions support network learning, we used node strength to identify hubs. Node strength is defined as the sum of all the connection weights to an ROI. Node strength was measured by averaging the subject functional connectivity matrices and then computing the sum of the connectivity of each node with all other nodes. We computed the node strength for the combined social and non-social brain network (averaging across the social and non-social tasks) and the social versus non-social brain network (subtracting the average non-social connectivity matrix from the average social connectivity matrix). We repeated this procedure for our 500 null model networks, and *z*-scored the node strength statistics. We identified hubs that had significantly greater node strength (FDR-corrected $p<0.05$) for transition trials than for non-transition trials across both the social network learning condition and the non-social control condition (domain-general hubs), hubs that had significantly greater node strength for the social network condition than for the non-social control condition (social hubs), and hubs that had significantly greater node strength for the non-social control condition than for the social network condition (non-social hubs).

After identifying hubs that were recruited during network learning, we next sought to determine which other brain regions were most strongly connected to those hubs. We began by computing the average strength of connectivity within and between sets of hubs in *a priori* regions of interest. In particular, we were interested in hippocampus, mPFC, IFG, and dlPFC based on previous work,[11] as well as TPJ and amygdala based on their roles in social processing[39]. We further broke this down into left and right portions of the following regions: hippocampus, TPJ, vmPFC, dmPFC, IFG, and dlPFC. We computed the average strength of connectivity within and between each set of ROIs. We first averaged the connectivity strength across all hubs within each of the *a priori* ROIs across both the social and non-social networks to obtain each ROI's within-region domain-general connectivity. Next, we averaged the connectivity between each pair of ROIs to compute the between-region domain-general connectivity. We also computed the within-region and between-region average connectivity for the social versus non-social connectivity matrix. We then repeated this procedure for our 500 null model networks, and *z*-scored the average ROI connectivity statistics for the domain-general analysis and the social versus non-social analysis.

**Cognitive Systems Involved in Network Learning**

We next examined whether the hubs identified based on node strength were clustered in particular cognitive systems. Using the Yeo 17-system parcellation which subdivides cognitive control, default mode, dorsal attention, ventral attention, somatomotor, limbic, and visual systems into 2-3 subsystems each[43,45], we counted the number of hubs that were located in each system. Because the hubs were defined based on the null models, we randomly permuted the system labels for each hub 500 times; this procedure allowed us to maintain the total number of hubs and the total number of nodes in each system. We then compared the actual number of hubs in each system to the distribution of hubs in each system from the 500 permutations to calculate $z$-scores and $p$-values for each system. These statistics thus represented the probability that at least that many hubs were located in each system compared to randomly assigned system labels.

We also computed the average connectivity within and between each set of systems. We first averaged the connectivity across all regions within each of the Yeo cognitive systems across both the social network condition and the non-social control condition to compute each system's within-system domain-general connectivity. Next, we averaged the connectivity between each pair of systems to compute the between-system domain-general connectivity. We also computed the within-system and between-system average connectivity for the social versus non-social connectivity matrix. We then repeated this procedure for our 500 null model networks, and z-scored the average system connectivity statistics for the domain-general analysis and the social versus non-social analysis.

**Association between Network Learning and Brain Hubs**

In addition to examining which brain regions operate as hubs supporting network learning, we were also interested in whether individual differences in hub connectivity might be associated with individual differences in learning. That is, are individuals with stronger connectivity in the hubs identified above also better at learning network structure? To address this question, we took the subject-level connectivity matrices and computed the average node strength for each set of hubs (domain-general hubs, social hubs, and non-social hubs) for each subject. We converted the node strength metric to $z$-scores for each subject by comparing the average node strength to the distribution of average node strengths based on the 500 null model networks for each subject. We then computed the Pearson correlation coefficient for the $z$-scored node strength in each set of hubs with the cross-cluster surprisal for each task condition. Adding head motion and task order as covariates did not alter the results that we report.

In addition to examining the relationship between node strength and network learning, we also examined whether average connectivity within each set of hubs was associated with network learning. To address this question, we computed the average connectivity within each set of hubs (domain-general hubs, social hubs, and non-social hubs) for each subject, converted each value to $z$-scores based on the non-parametric statistical approach described above. We then computed the Pearson correlation coefficient for the $z$-scored average connectivity with cross-cluster surprisal for each task. Adding head motion and task order as covariates did not alter the results that we report.

**Data Availability**
The data and code to reproduce all analyses and figures in this paper are available in Github repository [https://github.com/stompson/Tompson_Network_Learning_fMRI].

# Supplementary Materials

## Supplementary Method

In addition to examining connectivity between brain regions, we also examined task-related activation for transition versus non-transition trials for the social versus non-social network learning tasks. Neuroimaging data was preprocessed using nipype[69] implemented in python 2.7 using a combination of AFNI[71], FSL[72,73], and ANTs[74]. The functional data first underwent de-spiking to smooth outliers in each voxel using AFNI's 3dDespike. We next used FSL's MCFLIRT[75] to apply a rigid transformation to correct for head motion, and then we used FSL's Slicetimer to control for temporal differences in the order of the acquisition of the slices in each brain volume. Skull stripping was performed on the structural data using FSL's BET. The mean functional image was computed and bias-corrected using N4 Bias Field Correction implemented in ANTs, and then the skull-stripped structural image was bias-corrected. ANTs was used to compute the transformation parameters for the bias-corrected mean functional image to the bias-corrected and skull-stripped high resolution structural image. ANTs was used to warp the skull-stripped high-resolution structural image to the MNI152 2mm brain template provided by FSL. The transformation parameters from the ANTs functional to structural co-registration and the transformation parameters from the ANTs structural to MNI co-registration were used to warp the 4D functional image to the MNI152 2mm brain template. Finally, FSL's fslmaths was used to apply a 6mm Gaussian kernel to smooth the warped 4D functional image.

We then constructed a first-level general linear model (GLM) for each participant and each task with a boxcar function for transition versus non-transition trials convolved with a canonical hemodynamic response function. Additional regressors of no interest included six motion parameters derived from the motion correction preprocessing step. All first-level analyses steps were conducted with nistats in python 2.7. We next constructed a second-level GLM with an intercept modeling the trial type (transition versus non-transition), a regressor for network type (social versus non-social), and a regressor for subject (to account for the within-subjects design). We tested the effect of trial type using a one-sample t-test and the effect of network type using a paired-samples t-test. Results were thresholded using a whole-brain FDR-corrected $p<0.05$ and cluster-threshold of $k>5$. All second-level analyses steps were conducted with nistats in python 2.7.

## Supplementary Results

**Behavioral evidence for network learning**

In the main manuscript, we performed an initial assessment of the behavioral data that sought to extract evidence for network learning across both the social network learning task and the non-social control task. Here, we also ask whether there were important differences in behavior across the two tasks. We found that when the non-social control task was presented first, reaction times were significantly smaller than when the social network task was presented first ($B=0.026$, $SE=0.012$, $t(23.590)=2.111$, $p=0.046$). Moreover, we observed an interaction between node type and order, such that the cross-cluster surprisal effect was larger for participants who completed the non-social control task first ($B=0.035$, $SE=0.014$, $t(23.644)=2.579$, $p=0.017$). We did not hypothesize these effects *a priori*, and we suggest that further work should be done to ensure their reliability and replicability across other similar experiments.

**Activation for transition versus non-transition trials**

There was significantly greater activation for transition versus non-transition trials in visual cortex, parahippocampal gyrus, and dlPFC (see Supplementary Figure S1). There was significantly greater activation for non-transition trials versus transition trials in mPFC, vlPFC, striatum, PCC, inferior parietal lobe (IPL), and inferior temporal gyrus (ITG; see Supplementary Figure S1).

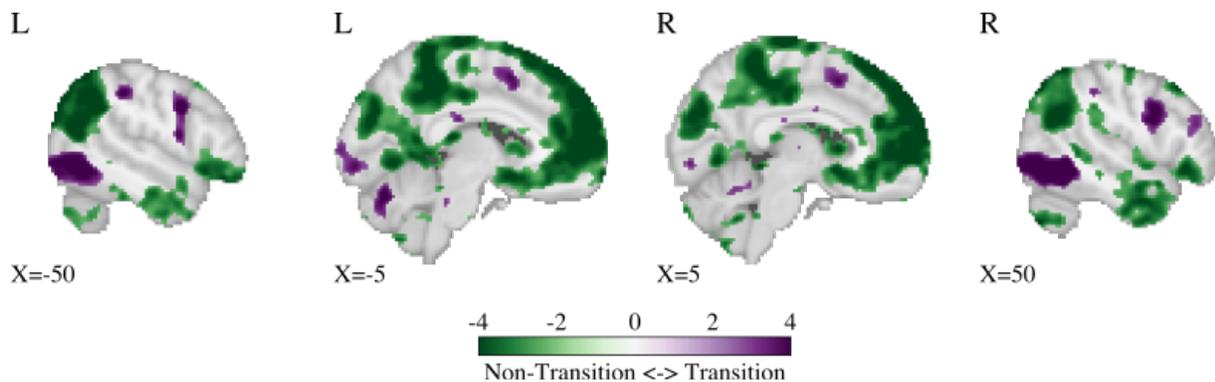

**Figure S1. Differences in brain activation for transition versus non-transition trials.** Whole-brain map showing regions with significantly different BOLD activation for transition (purple) versus non-transition (green) trials (whole-brain FDR corrected *p*<0.05).

**Activation for social versus non-social networks**

There were no brain regions with significantly different activation for social versus non-social tasks (FDR-corrected *p*<0.05).

# Supplementary Tables

Table S1. Node strength for both social and non-social networks

| Region | x,y,z[a] | Z[b] | k[c] |
|---|---|---|---|
| *Transition > Non-Transition Trials* | | | |
| R Dorsal Medial Prefrontal Cortex (dmPFC) | (12, 58, 36) | 3.291 | 3076 |
| L Superior Temporal Lobe | (-36, -26, 2) | 3.291 | 881 |
| R Inferior Temporal Lobe | (64, -26, -28) | 3.291 | 3201 |
| R Rectus Gyrus | (10, 12, -22) | 3.291 | 1703 |
| L Precentral Gyrus | (-12, -30, 62) | 3.291 | 135 |
| R Postcentral Gyrus | (68, -14, 16) | 3.291 | 1031 |
| R Postcentral Gyrus | (50, -16, 60) | 3.291 | 1238 |
| L Hippocampus | (-6, -14, -16) | 3.291 | 2636 |
| R Hippocampus | (26, -20, -15) | 3.291 | *** |
| L Paracentral Lobule | (0, -36, 54) | 2.878 | 214 |
| | | | |
| *Non-Transition > Transition Trials* | | | |
| R Middle Temporal Lobe | (68, -52, 0) | 3.291 | 9242 |
| L Middle Temporal Lobe | (-36, -66, 18) | 3.291 | 915 |
| L Precentral Gyrus | (-24, -16, 54) | 3.291 | 495 |
| L Inferior Parietal Lobe | (-30, -40, 44) | 3.291 | 284 |
| L Inferior Parietal Lobe | (-40, -66, 56) | 3.291 | 304 |
| L Occipital Lobe | (-20, -70, 36) | 3.291 | 374 |
| L Lingual Gyrus | (0, -82, -10) | 3.291 | 636 |
| L Insula | (-26, 22, -4) | 3.291 | 831 |
| L Superior Frontal Gyrus | (-12, -2, 64) | 3.291 | 460 |
| L Superior Frontal Gyrus | (-18, 56, -4) | 3.291 | 1745 |
| R Middle Frontal Gyrus | (50, 24, 38) | 3.291 | 383 |
| L Calcarine Gyrus | (-4, -60, 8) | 3.291 | 242 |
| R Temporoparietal Junction | (62, -36, 4) | 2.878 | 294 |
| L Inferior Parietal Lobe | (-52, -44, 40) | 2.652 | 199 |

*** Included in Left Hippocampus cluster

[a] Stereotactic coordinates from MNI atlas, in mm, left/right ($x$), anterior/posterior ($y$), superior/inferior ($z$), respectively, R = right, L = left

[b] Z-score, significant at FDR-correct $p<0.05$.

[c] Spatial extent in cluster size, threshold $\geq 5$ voxels.

Table S2. Hubs with significantly greater node strength for social than non-social condition

| Region | x,y,z[a] | Z[b] | k[c] |
|---|---|---|---|
| *Social > Non-Social Node Strength* | | | |
| R Temporoparietal Junction | (68, -46, 12) | 3.291 | 2443 |
| L Precentral Gyrus | (-30, 0, 50) | 3.291 | 2424 |
| R Postcentral Gyrus | (68, -10, 22) | 3.291 | 4657 |
| L Paracentral Lobule | (-6, -34, 68) | 3.291 | 783 |
| R Occipital Lobe | (30, -88, 34) | 3.291 | 6316 |
| L Middle Temporal Lobe | (-46, -10, -14) | 2.878 | 507 |
| L Inferior Temporal Lobe | (-22, -4, -44) | 2.878 | 604 |
| R Parahippocampal Gyrus | (38, -36, -18) | 2.878 | 348 |
| L Middle Cingulate | (-2, -32, 46) | 2.878 | 275 |

[a] Stereotactic coordinates from MNI atlas, in mm, left/right (*x*), anterior/posterior (*y*), superior/inferior (*z*), respectively, R = right, L = left

[b] Z-score, significant at FDR-correct p<0.05.

[c] Spatial extent in cluster size, threshold ≥ 5 voxels.

Table S3. Hubs with significantly greater node strength for non-social than social condition

| Region | x,y,z[a] | Z[b] | k[c] |
|---|---|---|---|
| *Non-Social > Social Node Strength* | | | |
| R Dorsal Medial Prefrontal Cortex (dmPFC) | (30, 62, 16) | 3.291 | 7036 |
| R Inferior Temporal Lobe | (70, -46, -4) | 3.291 | 452 |
| L Inferior Temporal Lobe | (-52, -46, -26) | 3.291 | 542 |
| R Inferior Parietal Lobe (IPL) | (64, -52, 32) | 3.291 | 3673 |
| R Occipital Lobe | (48, -84, -4) | 3.291 | 381 |
| L Occipital Lobe | (-14, -106, 8) | 3.291 | 2002 |
| L Occipital Lobe | (-20, -70, 36) | 3.291 | 408 |
| L Superior Frontal Gyrus | (-18, 56, -4) | 3.291 | 709 |
| R Inferior Frontal Gyrus (IFG) | (56, 36, 2) | 3.291 | 282 |
| R Inferior Frontal Gyrus (IFG) | (44, 14, -10) | 3.291 | 996 |
| L Inferior Frontal Gyrus (IFG) | (-22, 16, -16) | 3.291 | 1452 |
| R Inferior Frontal Gyrus (IFG) | (64, 10, 16) | 3.291 | 903 |
| R Middle Cingulate | (16, -44, 32) | 3.291 | 343 |
| R Rectus Gyrus | (10, 12, -22) | 2.652 | 309 |

[a] Stereotactic coordinates from MNI atlas, in mm, left/right (*x*), anterior/posterior (*y*), superior/inferior (*z*), respectively, R = right, L = left

[b] Z-score, significant at FDR-correct p<0.05.

[c] Spatial extent in cluster size, threshold ≥ 5 voxels.

Table S4. Brain regions with significantly greater hippocampal connectivity during the social network learning task

| Region | x,y,z[a] | Z[b] | k[c] |
|---|---|---|---|
| *Connectivity with Left Hippocampus* | | | |
| L Middle Cingulate | (0, -28, 30) | 3.291 | 271 |
| R Inferior Temporal Lobe | (64, -54, -16) | 3.291 | 1189 |
| R Temporoparietal Junction | (68, -44, 16) | 3.291 | 3672 |
| R Rolandic Operculum | (66, -4, 6) | 3.291 | 833 |
| L Precuneus | (0, -74, 30) | 3.291 | 550 |
| L Postcentral Gyrus | (-32, -20, 42) | 3.291 | 3313 |
| L Occipital Lobe | (-32, -94, -4) | 3.291 | 390 |
| L Lingual Gyrus | (-4, -66, -4) | 3.291 | 261 |
| R Insula | (48, -14, -6) | 3.291 | 261 |
| L Insula | (-30, 8, 10) | 3.291 | 1471 |
| R Middle Frontal Gyrus | (52, 40, 18) | 3.291 | 1544 |
| L Middle Cingulate | (-2, -32, 46) | 3.291 | 275 |
| R Ventral Striatum | (14, 16, -8) | 2.878 | 83 |
| L Temporoparietal Junction | (-46, -32, 0) | 2.878 | 563 |
| L Fusiform Gyrus | (-24, -62, -16) | 2.878 | 309 |
| R Cuneus | (26, -96, 20) | 2.878 | 469 |
| R Middle Frontal Gyrus | (54, 10, 40) | 2.652 | 521 |
| R Middle Frontal Gyrus | (34, -2, 52) | 2.652 | 304 |
| R Precentral Gyrus | (32, -30, 66) | 2.512 | 138 |
| L Postcentral Gyrus | (-20, -38, 60) | 2.512 | 269 |
| | | | |
| *Connectivity with Right Hippocampus* | | | |
| R Ventral Striatum | (14, 16, -8) | 3.291 | 83 |
| L Middle Temporal Pole | (-16, 0, -28) | 3.291 | 1003 |
| R Inferior Temporal Lobe | (70, -46, -4) | 3.291 | 452 |
| L Supplementary Motor Area | (0, -4, 68) | 3.291 | 206 |
| L Rolandic Operculum | (-42, 0, 6) | 3.291 | 259 |
| L Precentral Gyrus | (-36, -16, 38) | 3.291 | 567 |
| R Postcentral Gyrus | (64, -12, 40) | 3.291 | 636 |
| L Inferior Parietal Lobe (IPL) | (-40, -66, 56) | 3.291 | 304 |
| L Inferior Parietal Lobe (IPL) | (-48, -24, 38) | 3.291 | 1283 |
| L Occipital Lobe | (-10, -100, -18) | 3.291 | 2604 |
| R Middle Frontal Gyrus | (52, 40, 18) | 3.291 | 1553 |
| L Temporoparietal Junction | (-46, -32, 0) | 2.878 | 320 |

[a] Stereotactic coordinates from MNI atlas, in mm, left/right (*x*), anterior/posterior (*y*), superior/inferior (*z*), respectively, R = right, L = left

[b] Z-score, significant at FDR-correct p<0.05.
[c] Spatial extent in cluster size, threshold ≥ 5 voxels.

Table S5. Brain regions with significantly greater hippocampal connectivity during the non-social control task

| Region | x,y,z[a] | Z[b] | k[c] |
|---|---|---|---|
| *Connectivity with Left Hippocampus* | | | |
| L Middle Temporal Lobe | (-52, -24, -20) | 3.291 | 438 |
| L Supplementary Motor Area | (0, 2, 62) | 3.291 | 9103 |
| L Precuneus | (0, -62, 58) | 3.291 | 557 |
| L Superior Parietal Lobe | (-12, -84, 46) | 3.291 | 2405 |
| L Paracentral Lobule | (0, -40, 64) | 3.291 | 562 |
| R Dorsal Medial Prefrontal Cortex | (12, 58, 36) | 2.878 | 301 |
| L Middle Temporal Lobe | (-42, -60, 2) | 2.878 | 956 |
| R Inferior Temporal Lobe | (64, -20, -30) | 2.878 | 1159 |
| R Inferior Frontal Gyrus | (56, 30, -6) | 2.878 | 840 |
| | | | |
| *Connectivity with Right Hippocampus* | | | |
| L Middle Temporal Lobe | (-42, -60, 2) | 3.291 | 565 |
| R Inferior Temporal Lobe | (64, -20, -30) | 3.291 | 1460 |
| R Rectus Gyrus | (16, 44, -24) | 3.291 | 1467 |
| L Precuneus | (0, -72, 50) | 3.291 | 2890 |
| R Lingual Gyrus | (26, -54, 0) | 3.291 | 1617 |
| L Inferior Frontal Gyrus | (-22, 16, -16) | 3.291 | 1114 |
| R Temporoparietal Junction | (68, -46, 14) | 2.878 | 195 |
| R Superior Parietal Lobe | (30, -68, 60) | 2.878 | 368 |
| L Posterior Cingulate Cortex | (-10, -48, -6) | 2.878 | 118 |
| R Insula | (50, -16, 2) | 2.878 | 157 |
| L Anterior Cingulate Cortex | (0, -4, 32) | 2.878 | 143 |

[a] Stereotactic coordinates from MNI atlas, in mm, left/right (*x*), anterior/posterior (*y*), superior/inferior (*z*), respectively, R = right, L = left

[b] Z-score, significant at FDR-correct p<0.05.

[c] Spatial extent in cluster size, threshold ≥ 5 voxels.